\documentclass[12pt]{article}
\usepackage{amsmath}
\usepackage{amsfonts}
\usepackage{amssymb}
\usepackage{amsthm}
\usepackage{stackrel}

\usepackage{color}

\usepackage{epsfig}

\usepackage{fancyhdr}

\usepackage{graphics}

\def\binom#1#2{{#1}\choose{#2}}
\newcommand{\R}{{\rm I}\kern-0.18em{\rm R}}
\newcommand{\1}{{\rm 1}\kern-0.25em{\rm I}}
\newcommand{\E}{{\rm I}\kern-0.18em{\rm E}}
\newcommand{\p}{{\rm I}\kern-0.18em{\rm P}}
\makeatletter

\title{Number of clusters, deconvolution and classical problem of moments}
\author{Lev B. Klebanov\footnote{Department of Probability and Mathematical Statistics, MFF, Charles University, Prague 18675, Czech Republic.}, Zeev Volkovich\footnote{Software Engineering Department, ORT Braude College of Engineering, Karmiel 21982, Israel.}}
\date{}
\begin{document}
\maketitle

\begin{abstract}
In the paper there is given a connection between one special case of cluster analysis, deconvolution problem, and classical moment problem. Namely, the methods used there are applied to solve deconvolution problem for the case of one known distribution and another one concentrated in unknown finite number of points. These results can be applied to estimate a number of clusters for the case of scale or location mixture of identical distributions.

keywords: number of clusters, deconvolution problem, classical moment problem, scale and location mixtures. 
\end{abstract}

\section{Introduction}\label{sec1}
\setcounter{equation}{0} 

We study deconvolution problem for two cases. The first variant is to estimate one of the multiplicative convolution components when another component is known. The convolution itself is observable, and unknown component is concentrated at finite number of points. The second variant concerns the case of additive convolution.

Let us explain our problem and method on one simple example. Suppose that $Z$ is a random variable having standard normal distribution. Let $Y$ be a random variable, independent of $Z$ and taking $k$ unknown positive values $\sigma_1, \sigma_2, \ldots ,\sigma_k$ with probabilities $p_j=\p\{Y=\sigma_j\}$, $j=1, \ldots ,k$. The number $k$ is supposed to be fixed, but unknown. Denote $X=Y\cdot Z$, and suppose that we have $n$ observations of $X$, that is $n$ independent identically distributed (i.i.d.) with $X$ random variables $X_1, \ldots , X_n$. Our aim is to estimate the number $k$ and the parameters $\sigma_j, \; j=1, \ldots ,k$ on the base of observations $\{X_i, \; i=1, \ldots ,n \}$. It is possible to interpret the distribution of $X$ as normal laws scale mixture. Each normal distribution with variance $\sigma_j^2$ may be considered as a cluster. The number of clusters is $k$, and the weight of $j$th cluster is $p_j$. We have to estimate numbers $k$ and $p_j$, ($j=1, \ldots ,k$). In view of independence of $Z$ and $Y$, the moments $\mu_j(X) = \E X^j$, ($j=1, 2,\ldots $)  are products of corresponding moments $\mu_j(Z)$ and $\mu_j(Y)$: 
\[\mu_j(X)=\mu_j(Z)\mu_j(Y), \; j=1,2,\ldots .\]
Because the distribution of $Z$ is known, we may estimate the moments $\mu_j(Y)$ using empirical moments instead of $\mu_j(X)$. Without loss of generality, we may change random variable $Y$ by another variate (denoted by $Y$ again) which has symmetric distribution taking the values $\pm \sigma_j$ with probabilities $p_j/2$ each, so that $\mu_{2*j-1}(Y)=0$ for $j=1, 2, \ldots$. Now we may apply the theory for classical problem of moments (see, for example, \cite{Akh}). The condition of positiveness of thruncated sequence of moments gives us an estimate for $k$, while the parameters $p_j$ may be estimated on the base of corresponding orthogonal polynomials.

The problem mentioned above may be considered as deconvolution problem. To see this it is sufficient to pass from random variables $X,Y,Z$ to logarithms of their absolute values. It allows to pass from multiplicative mixture to additive one. The problem takes now the foloowing form. It is necessary to estimate one component of its convolution with known distribution basing on observation of the convolution itself. It is so-called deconvolution problem.

\section{Procedure for a solution of deconvolution problem in the case of scale mixture}\label{sec2}
\setcounter{equation}{0} 

Here we give detailed procedure of deconvolution problem for the case of scale mixtures. We use notations of Section \ref{sec1}, but the distribution of $Z$ is not supposed to be Gaussian. However, we will impose some restrictions on the distribution.

\textsl{\textbf {Assumption A}. The distribution of random variable $Z$ possesses of all moments and can be recovered from them in unique way}.

\textsl{\textbf {Assumption B}. All moments of random variable $Z$ are non-zero}.

Let $X=Z\cdot Y$ be a product of independent random variables, where the distribution of $Z$ is known and satisfies Assumptions A and B. $Y$ is a random variable, independent of $Z$ and taking $k$ unknown positive values $\sigma_1, \sigma_2, \ldots ,\sigma_k$ with probabilities $p_j=\p\{Y=\sigma_j\}$, $j=1, \ldots ,k$. The number $k$ is supposed to be fixed, but unknown. Basing on a random sample $X_1, \ldots , X_n$ from the distribution of $X$ we have to estimate $k$ and $\sigma_1, \ldots , \sigma_k$.

\textsl{Suppose that both \textbf{Assumptions A and B} hold}. Denote by $m_j(n) = \frac{1}{n}\sum_{i=1}^{n}X_i^j$, $j=1,2, \ldots$ - empirical moments of $X$. Then, the value
\begin{equation}\label{eq1}
\mu_j^*(n)=m_j/\mu_j(Z), \;\; j=0,1,2,\ldots 
\end{equation}
may be used as consistent estimator of $\mu_j(Y)$. Because random variable $Y$ is concentrated in finite number $k$ of points, then 
\begin{equation}\label{eq2}
\mathcal{D}_s=\det\left(\begin{matrix}
\mu_0(Y) & \mu_1(Y) & ...&\mu_s(Y)\\
\mu_1(Y) & \mu_2(Y) & ...& \mu_{s+1}(Y)\\
...& ...& ...& ...\\
\mu_s(Y) & \mu_{s+1}(Y)&...&\mu_{2s}(Y)\\
\end{matrix}
\right)>0,
\end{equation}
for all $s<k$, and determinant (\ref{eq2}) equals to 0 for $s=k$ and, therefore, for all $s>k$ (see, \cite{KN}). Therefore, we propose to use the following procedure for estimation of the number $k$. Consider determinants
\begin{equation}\label{eq3}
\mathcal{D}_s^* =\det\left(\begin{matrix}
1 & \mu_1^*(n) & ...&\mu_s^*(n)\\
\mu_1^*(n) & \mu_2^*(n) & ...& \mu_{s+1}^*(n)\\
...& ...& ...& ...\\
\mu_s^*(n) & \mu_{s+1}^*(n)&...&\mu_{2s}^*(n)\\
\end{matrix}
\right), 
\end{equation}
for the first value $s$ for which $\mathcal{D}_s^* \leq 0$. This value denote by $k^*=k^*(n)$, and use it as an estimator for $k$. $k^*(n)$ is a consistent estimator of $k$ because (\ref{eq1}) are consistent estimators for the moments of random variable $Y$. In other words, $k^*(n) \to k$ as $n \to \infty$ in probability. 

Let us make some remarks about this procedure:
\begin{enumerate}
\item Estimators $\mu_j^*(n)$ of $\mu_j(Y)$ are not moments themselves. Therefore, the determinants $\mathcal{D}_s^*$ are not obligated to be non-negative for all $s$. 
\item If the number $k$ is not small, then the order of first zero determinant (\ref{eq2}) is $2k$, which is also not small, and its calculation is not simple. We need high accuracy of calculations and, therefore, high precision of moments estimation, and, consequently, large number of needed observations $n$.
\item It is known, that the accuracy of estimators of the moments of high order is rather low. Therefore, we have one more argument, that the number of observations $n$ has to be very large for not small values of the components  number $k$.
\end{enumerate}

The problem of estimating the number of clusters is very interesting by itself, and therefore we provide simulation study of the quality of estimator $k^*$. The problem of estimating $\sigma_1,\sigma_2, \ldots , \sigma_k$ will be considered later.

We simulated samples of different sample size from mixtures of exponential distributions. On their basis we estimated the number $k$ of components in the mixtures. The results are given below.
\begin{itemize}
\item Simulation of samples from mixture of $k=2$ exponential distributions with scale parameters $\lambda_1=1$ and $\lambda_2=1/3$ (mean values are 1 and 3 correspondingly) with equal weights. Sample size used $n=100$. There were 5000 simulations. The estimator $k^*=k^*(100)$ had the following parameters: mean value $\E (k^*) = 1.951$; median $me(k^*)=2$; standard deviation $\sigma(k^*)=0.236$; percent of correct defined values $pr(k^*)=94.18\%$. The results here seem to be rather good.
\item  Simulation of samples from mixture of $k=3$ exponential distributions with scale parameters $\lambda_1=1$, $\lambda_2=3$ and $\lambda_2=5$ (mean values are 1,3 and 5) with equal weights. Sample size used $n=100$. There were 5000 simulations. The estimator $k^*=k^*(100)$ had the following parameters: mean value $\E (k^*) = 1.98$; median $me(k^*)=2$; standard deviation $\sigma(k^*)=0.182$; percent of correctly defined values $pr(k^*)=1.14\%$. The results seem to be very bad. However, it is clear, that the sample size is too small for such problem. 
\item Now the same situation as before, but with different sample sizes. We will see that: the mean value of $k^*$ becomes closer to $k=3$ with growing $n$, and percent of correctly defined values of $k$ grows with growing $n$, too. 
\subitem a) $n=1,000$; $\E (k^*) = 2.288$; $me(k^*)=2$; $\sigma(k^*)=0.455$; $pr(k^*)=29.22\%$.
\subitem b) $n=2,000$; $\E (k^*) = 2.401$; $me(k^*)=2$; $\sigma(k^*)=0.49$; $pr(k^*)=40.12\%$.
\subitem c) $n=3,000$; $\E (k^*) = 2.456$; $me(k^*)=2$; $\sigma(k^*)=0.498$; $pr(k^*)=45.58\%$.
\subitem d) $n=4,000$; $\E (k^*) = 2.495$; $me(k^*)=2$; $\sigma(k^*)=0.50$; $pr(k^*)=49.48\%$.
\subitem e) $n=5,000$; $\E (k^*) = 2.535$; $me(k^*)=3$; $\sigma(k^*)=0.499$; $pr(k^*)=53.44\%$.
\subitem f) $n=6,000$; $\E (k^*) = 2.542$; $me(k^*)=3$; $\sigma(k^*)=0.499$; $pr(k^*)=54.16\%$.
\subitem g) $n=7,000$; $\E (k^*) = 2.576$; $me(k^*)=3$; $\sigma(k^*)=0.496$; $pr(k^*)=57.36\%$.
\subitem h) $n=8,000$; $\E (k^*) = 2.614$; $me(k^*)=3$; $\sigma(k^*)=0.489$; $pr(k^*)=61.16\%$.
\subitem i) $n=9,000$; $\E (k^*) = 2.620$; $me(k^*)=3$; $\sigma(k^*)=0.488$; $pr(k^*)=61.7\%$.
\subitem j) $n=10,000$; $\E (k^*) = 2.651$; $me(k^*)=3$; $\sigma(k^*)=0.481$; $pr(k^*)=64.72\%$.

For $n=30,000$ we have  $\E (k^*) = 2.859$; $me(k^*)=3$; $\sigma(k^*)=0.422$; $pr(k^*)=80.22\%$.

\item  Simulation of samples from mixture of $k=4$ exponential distributions with scale parameters $\lambda_1=1$, $\lambda_2=1/3$, $\lambda_2=1/5$ and $\lambda_2=1/7$ shows that one needs form 100,000 to 1,000,000 observations to get the results similar to that for $k=3$. We omit the exact numbers.
\end{itemize}

From all above we can see, that to make sure there exist one or two clusters, it is enough to have hundreds observations. To make sure there are three clusters we need some thousands observations. For four components there is necessarily to have hundreds of thousands or just millions of observations. The reasons for that are mentioned in 1.--3. However, there is another, more essential, reason consisting in the fact that many-components system can be often approximated rather good by three-components distributions. To see the difference between that one needs much more observations than for small-components cases. Of course, the number of observations needed for a good agreement between estimator and reality depends of the $Y$ distribution structure.

Let us now return to the problem of estimating the parameters $\sigma_1, \sigma_2, \ldots , \sigma_k$. Define a sequence of polynomials
\begin{equation}\label{eq4}
P_s^*(\lambda)=P_s^*(\lambda;n)=\frac{1}{\sqrt{\mathcal{D}_{s-1}^*\mathcal{D}_{s}^*}}\det\left(\begin{matrix}
1 & \mu_1^*(n) & ...&\mu_s^*(n)\\
\mu_1^*(n) & \mu_2^*(n) & ...& \mu_{s+1}^*(n)\\
...& ...& ...& ...\\
\mu_{s-1}^*(n) & \mu_{s}^*(n)&...&\mu_{2s-1}^*(n)\\
1 & \lambda & ... & \lambda^s\\
\end{matrix}
\right), 
\end{equation}
where $\mathcal{D}^*_s$ is defined by (\ref{eq3}). It is clear, that (\ref{eq4}) gives us a consistent estimator for the following polynomial
\begin{equation}\label{eq5}
P_s(\lambda)=\frac{1}{\sqrt{\mathcal{D}_{s-1}\mathcal{D}_{s}}}\det\left(\begin{matrix}
1 & \mu_1 & ...&\mu_s\\
\mu_1 & \mu_2 & ...& \mu_{s+1}\\
...& ...& ...& ...\\
\mu_{s-1} & \mu_{s}&...&\mu_{2s-1}\\
1 & \lambda & ... & \lambda^s\\
\end{matrix}
\right),
\end{equation}
$s=1,2, \ldots , k-1$. We cannot use this expression for $s=k$ because $\mathcal{D}_k =0$. Therefore, we omit the multiplier $1/\sqrt{\mathcal{D}_{s-1}\mathcal{D}_{s}}$ for the case $s=k$. Put $P_0(\lambda)=1$, so that the polynomials $P_s(\lambda)$ are defined for $s=0,1, \ldots ,k$. We can define $P_s^*(\lambda)$ for $s=0$ and $s=k^*$ in a natural way. By solving equation
\begin{equation}\label{eq6}
P_{k*}(\lambda)= 0
\end{equation}
we find its roots $\lambda_1^*, \lambda_2^*, \ldots , \lambda_{k^*}^*$. These values are consistent estimators for the roots $\lambda_1, \lambda_2, \ldots ,\lambda_k$ of polynomial $P_k(\lambda)$. It is known \cite{Akh}, that the roots of $P_k(\lambda)$ are the points of growth of the distribution of random variable $Y$. However, as has been mentioned above, the values $\mu_j^*(n)$ are not moments of any distribution, and therefore determinant $\mathcal{D}_{k^*}$ may be negative. To finish the process of estimation of the $Y$ distribution we need to estimate the weights $\sigma_j$ at points $\lambda_j$, ($j=1,2, \ldots , k$). It is known (see \cite{Akh}), that
\begin{equation}\label{eq7}
\sigma_j = \frac{1}{\sum_{i=0}^{k-1}|P_{i}(\lambda_j)|^2},\;\; j=1, \ldots ,k.
\end{equation} 
From (\ref{eq7}) we see that
\begin{equation}\label{eq8}
\sigma_j^* = \frac{1}{\sum_{i=0}^{k^*-1}|P_{i}^*(\lambda_j^*)|^2},\;\; j=1, \ldots ,k^*
\end{equation}
are consistent estimators of $\sigma_j$.

Summarizing, we can give the complete procedure for estimating of the $Y$ distribution under \textsl{\textbf{Assumptions A and B}} as follows.
\begin{enumerate}
\item Calculate determinants (\ref{eq3}) until the first $s$ for which $\mathcal{D}_s^* \leq 0$. This value denote by $k^*=k^*(n)$, and use it as an estimator for $k$.
\item Calculate polynomials (\ref{eq4}) for $s=1, \ldots , k^*-1$ and $P_0^*$, $P_{k^*}^*(\lambda)$.
\item By solving equation (\ref{eq6})find its roots $\lambda_1^*, \lambda_2^*, \ldots , \lambda_{k^*}^*$. These roots are estimators for growth points of the $Y$ distribution. 
\item Values (\ref{eq8}) give us estimators for the weights of $Y$ distribution.
\item Finally, the estimator for $Y$ distribution is a measure, concentrated in $k^*$ points $\lambda_1^*, \lambda_2^*, \ldots , \lambda_{k^*}^*$ and giving them weights $\sigma_j^*$, ($j=1, \ldots ,  k^*$).
\end{enumerate}

Is it possible to avoid \textbf{Assumption B}? In general situation the answer is negative. The reason is that we cannot find corresponding moments of $Y$ in the case when $Z$ has some moments equal to zero. However, one can still get some information on $Y$ distribution. For example, we can change the moments of $X$ by its absolute moments. This will allow us to estimate consistently the distribution of its absolute value. The number of growth points for $|Y|$ is not less than a half of that for $Y$. The points of growth them selfs for $Y$ are concentrated at that of $|Y|$ or at points symmetric to them. Let us note, that for the case of $|Y|$ all points of growth have to be non-negative.

\section{Procedure for a solution of deconvolution problem in the case of location mixture}\label{sec3}
\setcounter{equation}{0} 

Here we give detailed procedure of deconvolution problem for the case of location mixtures. We use notations of Sections \ref{sec1}, \ref{sec2}. Although the case of location mixtures may be transformed to that of scale by passing to exponents of corresponding random variables, the direct calculations appears to work better, and this is the reason for separate consideration of the location case. 

So, we start with consideration of the scheme
\begin{equation}\label{eq9}
X=Z+Y,
\end{equation}
where $X$ is observable random variable, $Z$ and $Y$ are independent random variables, the distribution of $Z$ is known, while about that of $Y$ distribution is known only that it is concentrate in a finite (unknown) number of points. We supposed that the  \textsl{\textbf{Assumption A}} holds. From (\ref{eq9}) it follows that
\[ \mu_s(X) = \sum_{j=0}^s \,{\binom{s}{j}}\mu_j(Z)\, \mu_{s-j}(Y), \; s=1,2, \ldots .\]
From here it is easy to obtain a formula for recurrent calculation of estimators for the moments of $Y$:
\begin{equation}\label{eq10}
\mu_0^*(n)=1, \;\;\mu_s^*(n) = m_s - \sum_{j=1}^s {\binom{s}{j}}\mu_j(Z)\mu_{s-j}^*(n), \;\; s=1,2, \ldots
\end{equation}
The rest of procedure is the same as in Section \ref{sec2}.

\section{Simulations and Applications}
\setcounter{equation}{0} 

Here we give some simulations and applications to analysis of stocks prices and to currency exchange rate. 

\subsection{Simulations}
Here we give simulation results for both scale and location mixtures. 
\begin{itemize}
\item[I.] {\it In the case of scale mixture} we simulated $5,000$ samples of volume $n=1,000$ each from mixture of two ($k=2$) normal distributions $N(0,1)$ and $N(0,9)$ with equal weighs. The estimator $k^*$ took true value $2$ in $3,084$ cases. Mean value of estimators for the standard deviation (equal to 1 in general population) of the first component was $0.8065$. Its weight had mean value $0.4752$. For the second component of the mixture we had mean value of estimators for the standard deviation (3 in general population) equals to $2.9835$. Its weight was estimated as $0.5201$ in the mean. From these simulations we see that the estimator for mean of the first component is biased to the left. However, this estimator may be used as initial approximation for maximum likelihood estimation (or other) procedure.
\item[II.] {\it In the case of location mixture} we simulated $5,000$ samples of volume $n=1,000$ each from mixture of two ($k=2$) normal distributions $N(-1,1)$ and $N(1,1)$ with equal weighs. The estimator $k^*$ took true value $2$ in $2,896$ cases. Mean value of estimators for the mean value (equal to -1 in general population) of the first component was $-0.9957$. Its weight had mean value $0.5007$. For the second component of the mixture we had mean value of estimators for the mean value (1 in general population) equals to $0.9957$. Its weight was estimated as $0.4993$ in the mean.
\item[III.] Again, in the case of location mixture $5,000$ samples of volume $1,000$ each from the mixture of two Laplace distributions with unite standard deviation and mean values $-1$ and $1$ were simulated. The estimator $k^*$ took true value $2$ in $3,041$ cases. Mean value of estimators for the mean value (equal to -1 in general population) of the first component was $-0.9886$. Its weight had mean value $0.4991$. For the second component of the mixture we had mean value of estimators for the mean value (1 in general population) equals to $0.9875$. Its weight was estimated as $0.5009$ in the mean.
\end{itemize}

\subsection{Financial applications}

{\bf A.} Let us consider some data on stock prices. The date were obtained on the site: http://finance.yahoo.com/q/hp?a=03\&b=12\&c=1996\&d=02\&e=23\&f=2016\&g=d\&s=MSFT\%2C+\&ql=1.

\begin{figure}[h]
\centering
\hfil
\includegraphics[scale=0.6]{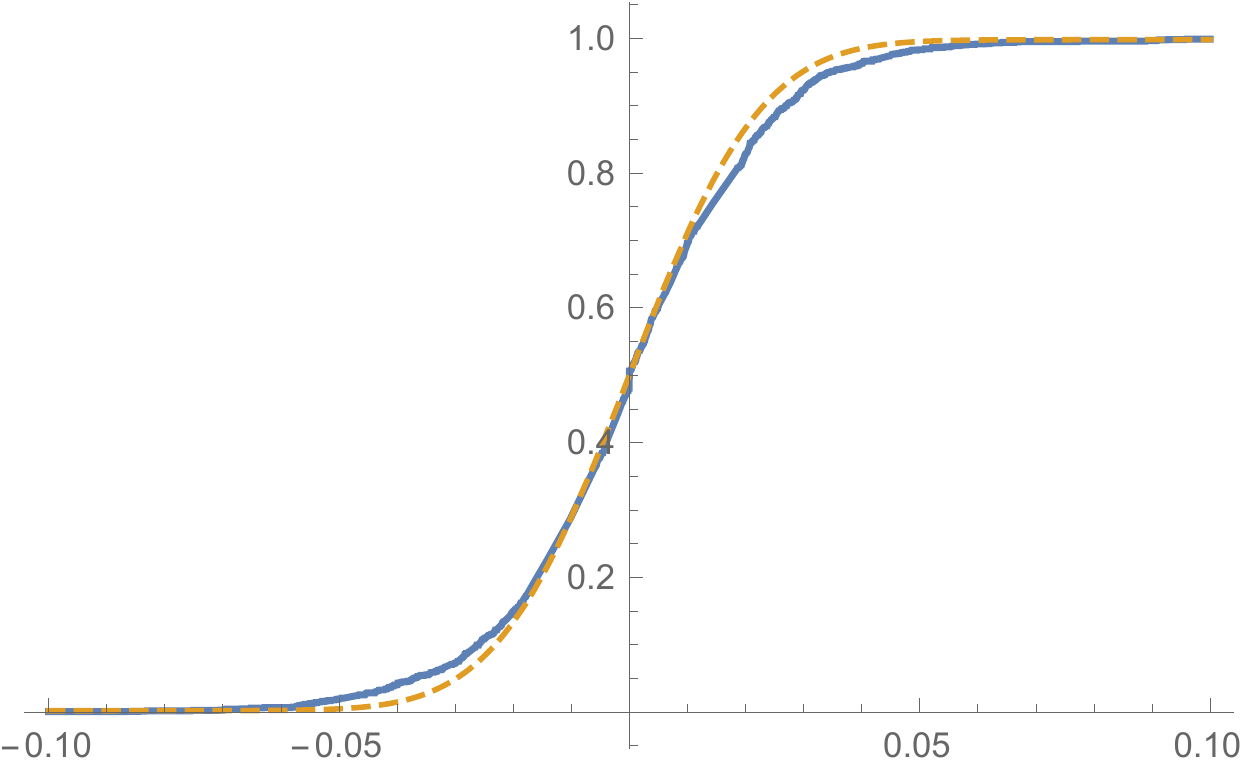}
\caption{Plot of the empirical distribution function (thick line) and model (dashed line)}\label{fig1}
\end{figure}

We analyze stock price changes of GazpromDE for the period from 13-11-2009 till 23-3-2016. Denote by $S_j$ the price of stock at the moment $j$, and introduce 
\[ X_j=\log(S_{j+1}/S_j),\;j=1, \ldots , n,   \]
where $n+1$ corresponds to the number of observed prices. In dataset under consideration it equals to $1,651$, so that $n=1,650$. Applying algorithm of Section \ref{sec2} we find $k^*=2$ for a mixture of normal distributions with parameters $(0,0.01788^2)$ and $(0, 0.301205^2)$. The mixture has weights $0.9948$ and $0.005215$. The agreement between empirical distribution and this model is not too good (see Figure \ref{fig1}).

\begin{figure}[h]
	\centering
	\hfil
	\includegraphics[scale=0.6]{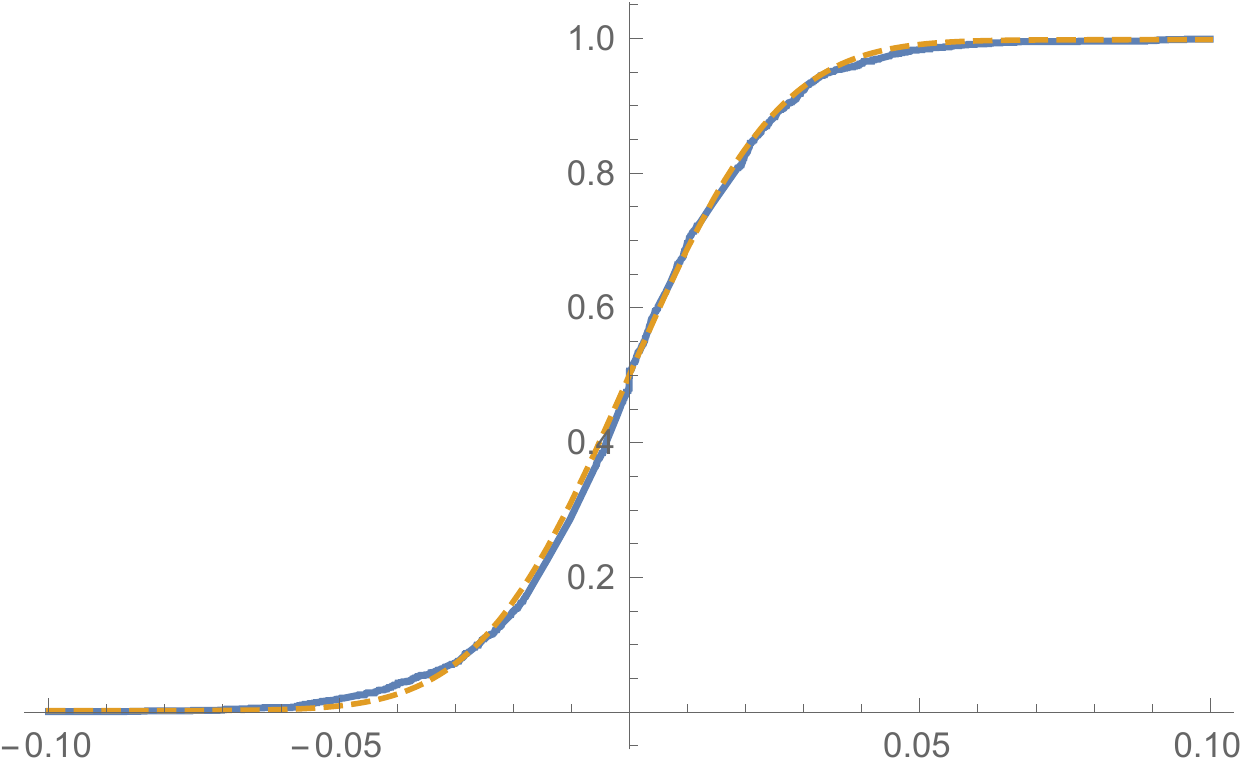}
	\caption{Plot of the empirical distribution function (thick line) and model after maximum likelihood correction (dashed line)}\label{fig2}
\end{figure}

Applying maximum likelihood method we find more precise values of normal components variances: $0.00036285$ for the first component and $0.1030$ for the second. The agreement between empirical data and corrected model is given on Figure \ref{fig2}.

We applied two statistical goodness-of-fit tests to verify agreement between empirical data and corrected model. P-Value for Anderson-Darling test is $0.06708$.
Cram\'{e}r-von Mises test has P-Value equal to $0.1476$. We see, that the both test did not reject the hypothesis that sample was drawn from the mixture of normal distributions with zero means.

\begin{figure}[h]
	\centering
	\hfil
	\includegraphics[scale=0.6]{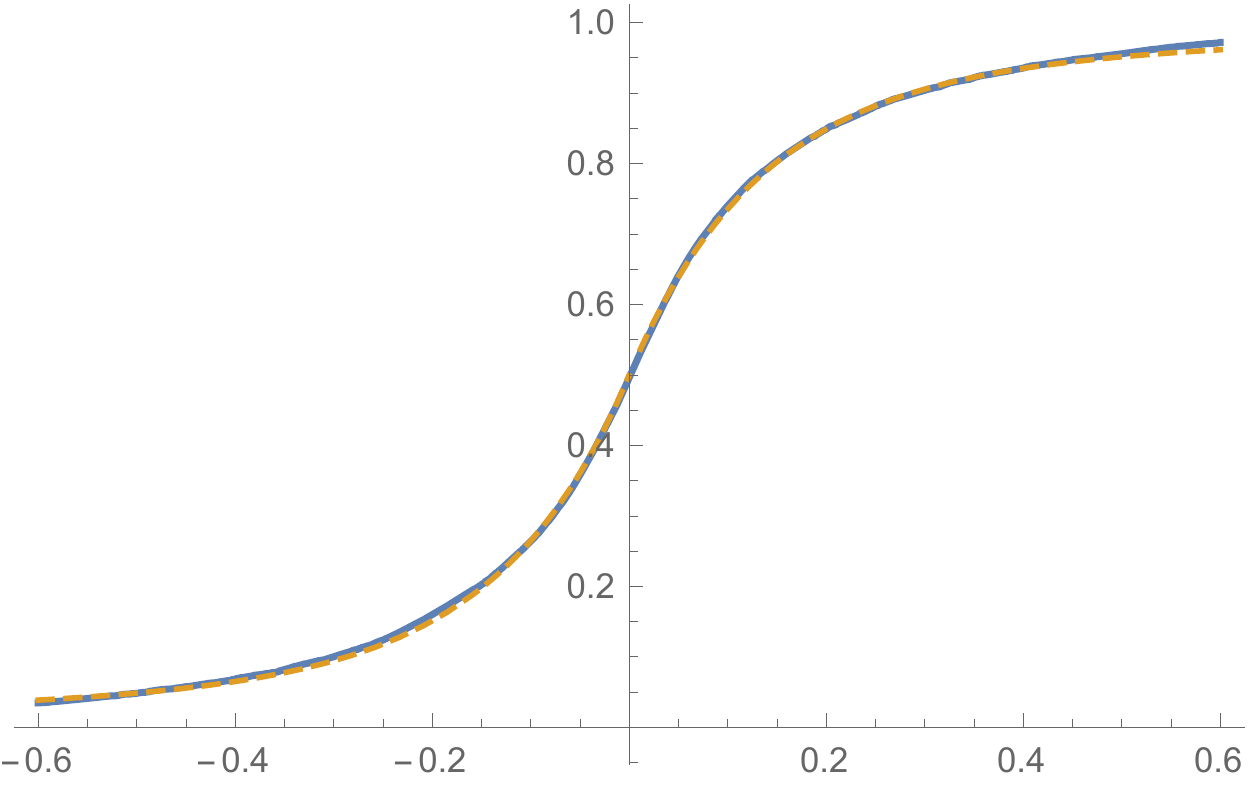}
	\caption{Plot of the empirical distribution function (thick line) and model after maximum likelihood correction (dashed line) for Dollar/RUB exchange rate }\label{fig3}
\end{figure}

{\bf B.} Now we consider exchange rate for US Dollar and Ruble of Russian Federation, the data for period from 11-12-2014 till 17-04-2015 with one minute interval. Denote $R_j$ the exchange rate at moment $j$. Introduce $\Delta_j =R_{j+1}-R_j$ and remove duplicate values of these differences. Remaining number od observations is $n=8,180$. After applying the algorithm of Section \ref{sec2} we find $k^* = 2$ for a mixture of Laplace distributions with zero means and scale parameters $0.1048$ and $0.5840$. Corresponding weights are $0.8438$ and $0.1562$. After applying maximum likelihood method we obtain new scale parameters of Laplace distribution. They are $0.1308$ and $0.7288$. Agreement between the model and empirical data is given on Figure \ref{fig3}.

We applied two statistical goodness-of-fit tests to verify agreement between empirical data and corrected model. P-Value for Cram\'{e}r-von Mises test equals to $0.3543$, and for Kolmogorov-Smirnov test it is $0.1434$. We see, that the both test did not reject the hypothesis that the sample was drawn from the mixture of Laplace distributions with zero means.

\subsection{Remarks on multidimensional case}

The case when $X=Y\cdot Z$ with multidimensional random vector $Z$ seems to be very interesting. However, one can change it to one-dimensional by taking projection on a direction $e$, that is by considering inner products $(X,e)=Y\cdot (X,e)$. Theoretically, it is necessary to consider projections on all vectors $e$ from unit sphere. However, if the vector $e$ is taken ``at random", one often (with probability 1) has this vector at ``general position", and it is enough to use only one projection. 

\section{Acknowledgment}
The work was partially supported by Grant GA\u{C}R 16-03708S.

\end{document}